\documentclass[showpacs,superbib,11pt,preprint,final,onecolumn,aps]{revtex4}

\usepackage{graphicx}
\usepackage{textcomp}
\usepackage{amsmath}

\begin{document}

\title{Current-induced forces: a simple derivation}

\author{Tchavdar N. Todorov} \author{Daniel Dundas}
\affiliation{Atomistic Simulation Centre, School of Mathematics and
  Physics, Queen's University Belfast, Belfast BT7 1NN, United
  Kingdom}

  \author{Jing-Tao
  \surname{L\"u}}  \affiliation{School of Physics,
   Huazhong University of Science and Technology,
   1037 Luoyu Road, Wuhan 430074, China}
\author{Mads \surname{Brandbyge}}
 \affiliation{Center for Nanostructured Graphene (CNG), DTU Nanotech,
  Department of Micro- and Nanotechnology, Technical University of
  Denmark, {\O}rsteds Plads, Build. 345E, DK-2800 Kongens Lyngby,
  Denmark}
\author{Per \surname{Hedeg{\aa}rd}}
\affiliation{Niels Bohr Institute, Nano-Science Center, University of
  Copenhagen, Universitetsparken 5, 2100 Copenhagen {\O}, Denmark}

\date{\today}

\begin{abstract}

We revisit the problem of forces on atoms under current in nanoscale conductors.
We derive and discuss the five principal kinds of force under steady-state conditions from a simple
standpoint that - with the help of background literature - should be accessible to physics
undergraduates. The discussion aims at combining methodology with an emphasis on the underlying
physics through examples. We discuss and compare two forces present only under current -
the non-conservative electron wind force and a Lorentz-like velocity-dependent force. It is shown 
that in metallic nanowires both display significant features at the wire surface, making it a
candidate for the nucleation of current-driven structural transformations and failure. Finally
we discuss the problem of force noise and the limitations of Ehrenfest dynamics.

\end{abstract}


\maketitle

\section{Introduction}

\begin{figure}[ht]
\begin{center}
\includegraphics[scale=0.45]{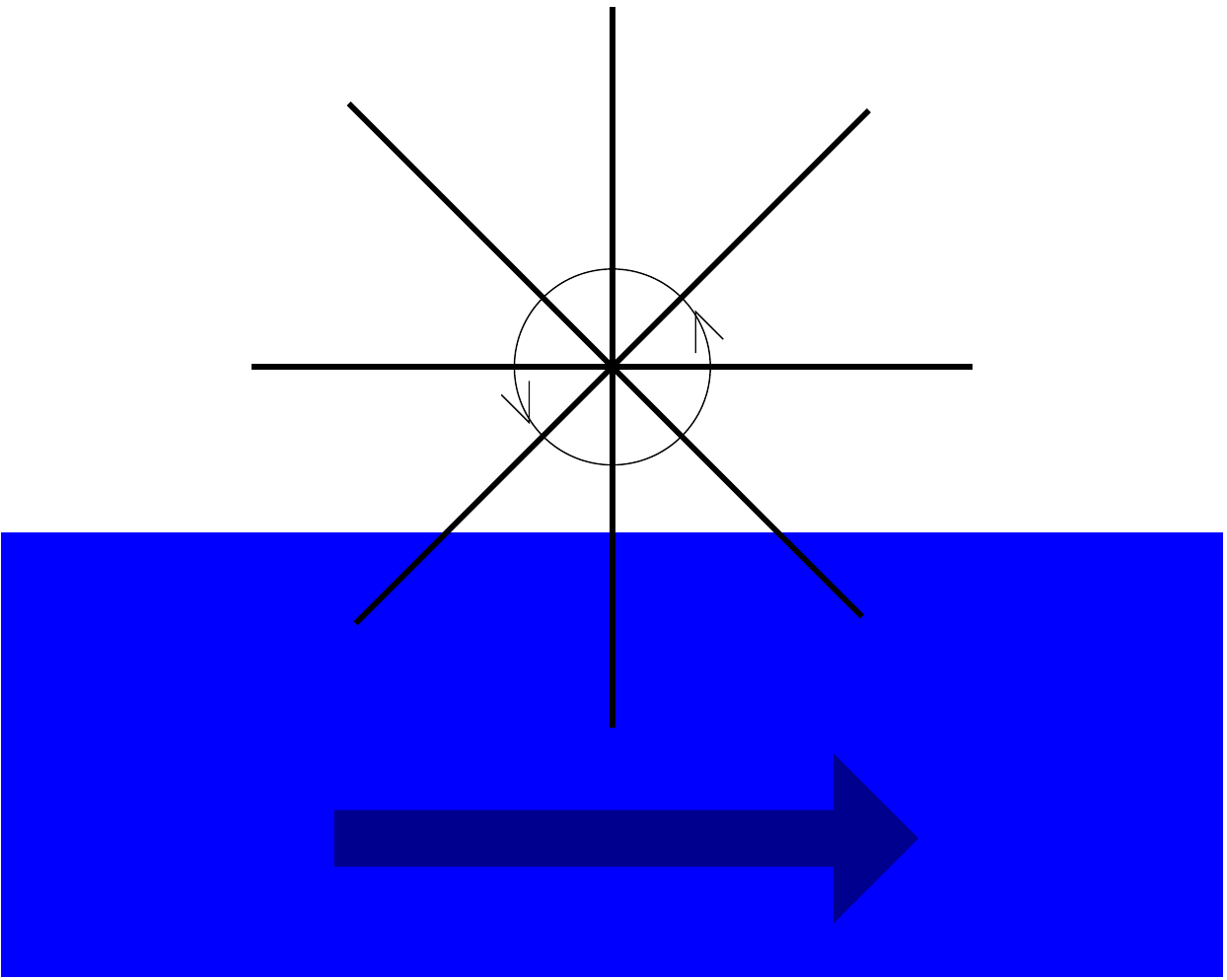}
\caption{\label{fig1} Schematic of a waterwheel.}
\end{center}
\end{figure}

We have all seen a waterwheel (Fig. \ref{fig1}) and have an intuitive
feel for how it works. Consider now an atom in a solid. Normally we think of atoms as
executing small-amplitude thermal vibrations about their equilibrium positions.
That is, we think of them as oscillators. In one dimension an oscillator can only
go back and forth. But in 2d or 3d it could run around a circle, provided its natural
frequencies along two independent directions are equal. Can then the flow of {\sl electrical
current} somehow be gauged to drive an atom around, in a manner akin to a waterwheel?
According to an argument by Sorbello \cite{sorbello} this should be possible.
When an atom - or more precisely an atomic nucleus - is immersed in a current,
incident electrons get scattered. There is a transfer of momentum between the flow
and the target, and a resultant force. This force is called the {\sl electron wind force}.
For an isolated scatterer in a free-electron metal it is given by
\begin{equation}
{\bf F}_{\rm w} = \sigma p_{F} {\bf j}
\label{fw}
\end{equation}
where $\sigma$ is the scattering cross section of the target \footnote{More precisely, $\sigma$ here
is the so-called transport cross section.}, $p_{F}$ is the Fermi momentum of
the electrons and ${\bf j}$ is the incident electron particle current density.

\begin{figure}[ht]
\begin{center}
\includegraphics[scale=0.4]{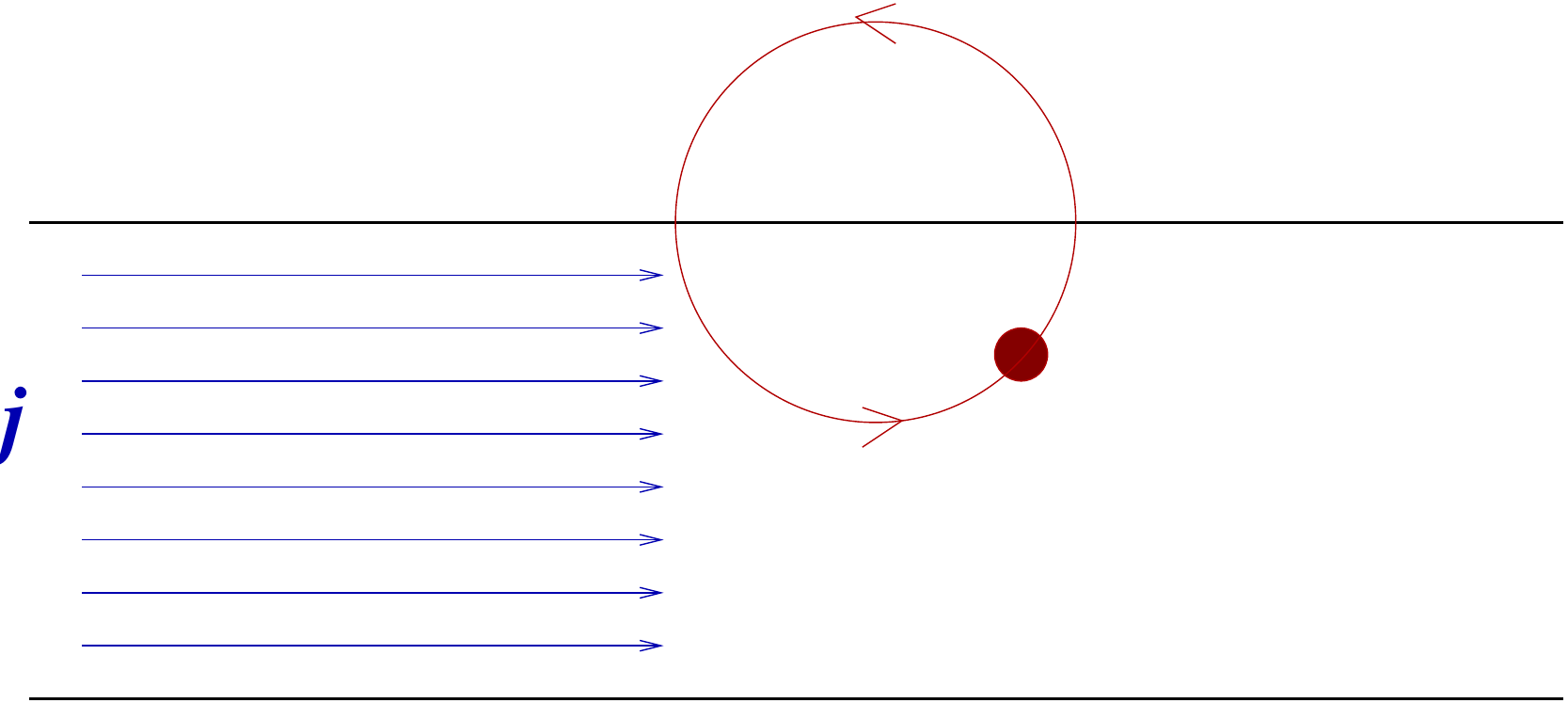}
\caption{\label{fig2} A scatterer in an electron current. When the scatterer is inside the metal wire,
the current density exerts a force on it - the electron wind force. Outside, there is no force.
Thus, the wind force does non-zero net work around the closed path shown, and can {\sl drive} the atom around.}
\end{center}
\end{figure}

Therefore if we design a closed path that goes through regions of different current density, the
net work done by this force need not be zero. One such path is shown in Fig. \ref{fig2}.
Then the wind force is {\sl non-conservative}, with the capacity to continually
pump energy into the nuclear motion by driving the nuclei around closed paths in configuration space.
This possibility began to attract attention in the context of atomic and molecular wires
about 10 years ago \cite{prl}, initially with arguments that appeared at odds with each other
\cite{pmb,prl}, the full resolution of which still requires further work \cite{molphys}.
After several years of further consideration the non-conservative character of forces on
atoms in nanowires was demonstrated microscopically, culminating in the design (through simulation)
of a single-atom ``waterwheel'' \cite{waterwheel}. The idea was soon extended to
waterwheels in abstract spaces spanned by generalized cooperative displacements \cite{mads1,mads2}.

These prospects raise a number of questions. Leaving aside the
idea of atomic-scale motors, non-conservative forces can have practical
repercussions for the {\sl stability} of atomic wires. Electromigration is a class of
phenomena \cite{sorbello} in which electrical current drives atomic diffusion
and rearrangements resulting in the mechanical failure of a conductor. Electromigration
is traditionally thought of as a thermally-assisted process. But kinetic energy gain
into ``waterwheel'' atomic motion under non-conservative forces could be an alternative
activation mechanism, and potentially a much more powerful one.

These problems are being investigated by a growing number of 
researchers \cite{bode1,bode2,mthomas,tnt1,tnt2,tnt3}
with further questions being uncovered all the time. However,
the physical - and to an extent mathematical - understanding of forces due to current remains
a challenging subject, for a simple reason. A real waterwheel is driven by a classical flow.
But in the atomic case it is the quantum electron fluid doing it, and the rotors themselves
(the nuclear subsystem) have to - or may have to - be considered quantum-mechanically,
at least for some aspects of the full problem.

Force on the other hand is fundamentally a classical notion.

Combining these partly but not wholly intuitive elements into a physical yet
rigorous understanding requires careful considerations. The above works have derived and discussed
forces under current from a variety of standpoints. They are fundamentally connected but vary
in technical difficulty and physical flavour.

The aim of the present discussion is to obtain these forces in a simple way which at the same time
retains the essential ingredients needed to capture the principal {\sl types} of forces - there are five -
along with a picture of the physics behind them. The methods involved should be
accessible to physics undergraduates and postgraduates in the physical sciences.

\section{Mean force}

The forces we are considering are fundamentally forces on atomic nuclei, or possibly ions.
Nevertheless we will occasionally speak of atoms and atomic motion, to help
visualize the problem.

\begin{figure}[ht]
\begin{center}
\includegraphics[scale=0.35]{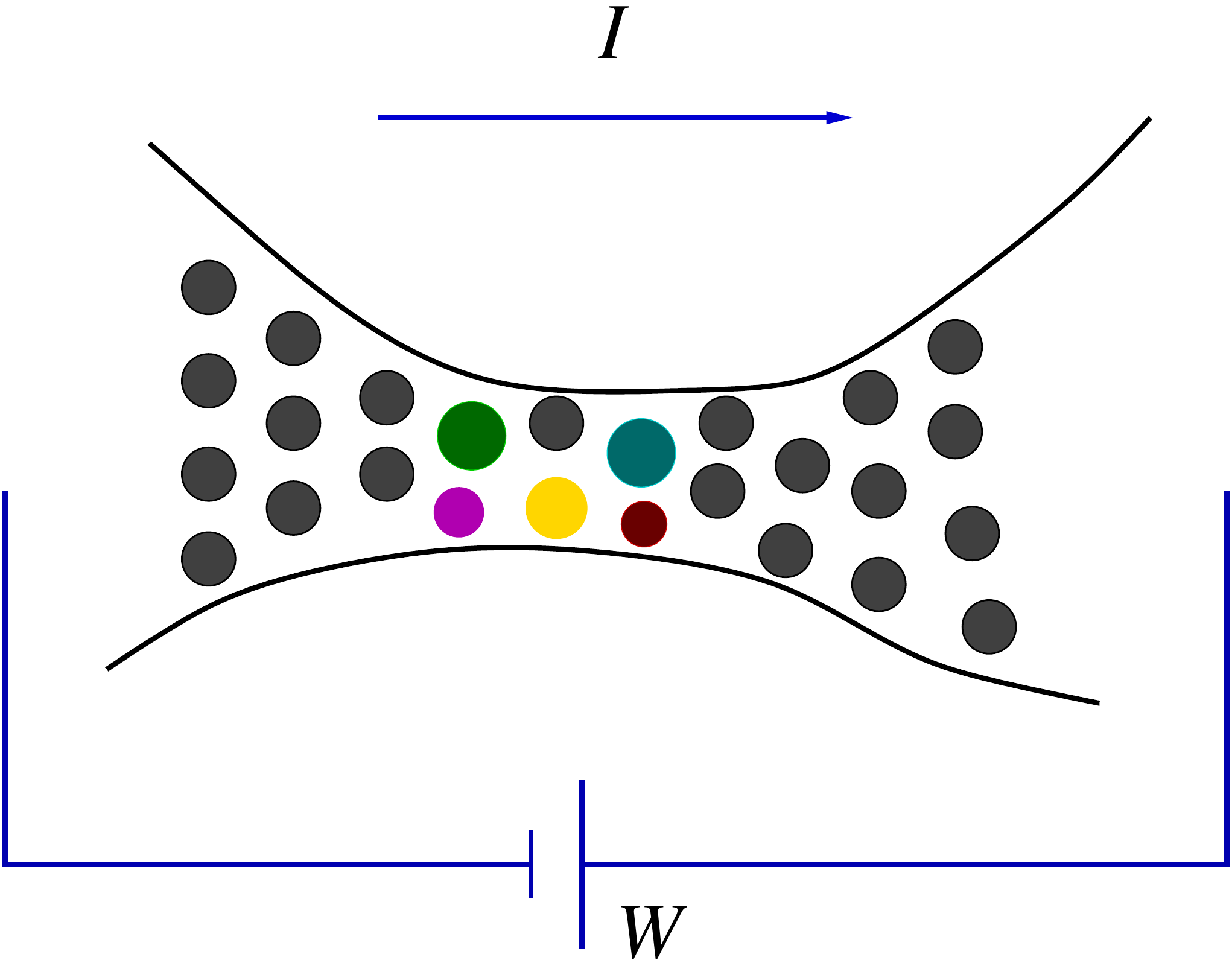}
\caption{\label{fig3} A conducting nanojunction. Circles represent atoms.}
\end{center}
\end{figure}

Our object is to find these forces under current flow in a generic atomic or
molecular wire \cite{agrait,galperin} as depicted in Fig. \ref{fig3}.
We will do so in steps, introducing briefly the Landauer picture of conduction in nanoscale systems
on the way.

\subsection{Formalism}

We will approach the problem by low-order time-dependent perturbation theory for the
density matrix (DM). The DM for a system with Hamiltonian
\begin{equation}
{\hat H}(t) = {\hat H}_{0} + {\hat V}(t)
\end{equation}
obeys the Liouville equation
\begin{equation}
{\rm i}\hbar \, {\dot {\hat \rho}}(t) = [{\hat H}(t),{\hat \rho}(t)]
\end{equation}
whose solution to first order in ${\hat V}(t)$ is
\begin{equation}
{\hat \rho}(t) = {\hat \rho}_{0}(t) + \frac{1}{{\rm i}\hbar}\,\int_{0}^{t}\,
[{\hat U}_{0}(t,\tau){\hat V}(\tau){\hat U}_{0}(\tau,t),{\hat \rho}_{0}(t)]\,d\tau
\label{1}
\end{equation}
where ${\hat U}_{0}(t,\tau) = {\rm e}^{-{\rm i}{\hat H}_{0}(t - \tau) / \hbar}$ and
${\hat \rho}_{0}(t) = {\hat U}_{0}(t,0){\hat \rho}(0){\hat U}_{0}(0,t)$.

Consider a system of quantum electrons and harmonic oscillators with
(for the moment, many-body) Hamiltonian
\begin{equation}
{\hat H} = \left[{\hat H}_{e} + \sum_{\beta}\,\left( \frac{{\hat P}_{\beta}^{2}}{2M_{\beta}} +
\frac{1}{2}M_{\beta}\omega_{\beta}^{2}\,{\hat X}_{\beta}^{2}\right)\right] - \sum_{\beta}\,{\hat
F}_{\beta}{\hat X}_{\beta} = {\hat H}_{0} + {\hat V}\,.
\label{heph}
\end{equation}
${\hat H}_{e}$ describes electrons in the absence of vibrations and the second term in square
brackets describes free vibrations (with $M_\beta$ a mass-like parameter). The two together form the
unperturbed Hamiltonian ${\hat H}_{0}$. ${\hat V} = - \sum_{\beta}\,{\hat F}_{\beta}{\hat X}_{\beta}$ 
is the electron-oscillator coupling, and will be treated as a perturbation.
For the moment the oscillators could represent generalized degrees of freedom, such as
unperturbed vibrational normal modes, or individual nuclear degrees of freedom viewed 
as Einstein oscillators. 

The construction of the coupling ${\hat F}_{\beta}$ from scratch is a 
difficult - and to an extent open - problem.
The systems of interest are infinite, aperiodic and open, with a continuum of 
electronic states. In addition, interatomic forces under current contain the key 
non-conservative component highlighted already. These factors require a reconsideration 
of the usual Born-Oppenheimer separation \cite{mthomas}. 

A possible physical picture of the {\sl unperturbed} dynamics
is to imagine that nuclei vibrate on a chosen known potential-energy surface (such as
the ground-state Born-Oppenheimer surface), while electrons remain in a 
current-carrying steady state for each geometry. One can then 
make a small-amplitude expansion in nuclear displacements about a chosen 
configuration. A natural choice of expansion point is the equilibrium 
geometry on the reference surface. However one may consider nearby points, so long 
as nuclei remain bound and a Hamiltonian quadratic in the displacements remains
appropriate, at least locally. The resultant coupling then has a dual role: 
to introduce the additional non-equilibrium current-induced forces, and 
to generate inelastic electron-phonon scattering. Explicit constructions of the 
coupling are discussed for example in Refs. \cite{disscomp,thomas}.

In general ${\hat F}_{\beta}$ will be the sum of a 1-body electronic operator 
and a scalar (both dependent on the expansion point). The electronic part 
typically is minus the gradient, with respect to nuclear displacements, of a 
(possibly screened) electron-nuclear interaction.

Alternatively we may think of (\ref{heph}) as a generic model
electron-phonon Hamiltonian, while noting that by construction ${\hat F}_{\beta}$
remains the gradient of the interaction with respect to the negative of the conjugate 
displacement. We will take this view here and will assume further that 
${\hat F}_{\beta}$ is purely a 1-body electronic quantity.

The resultant calculation is exact in the reference Hamiltonian, and is
exact also in the perturbation, in the limit of small-amplitude
atomic motion. This is important as the response coefficients that characterize
current-induced forces below describe that limit.

These questions are receiving renewed attention with fresh work on the 
many-body electron-nuclear Schr{\" o}dinger equation \cite{maitra1,maitra2,abedi}.

Our aim is to calculate the quantity
\begin{equation}
F_{\alpha}(t) = {\rm Tr}\{{\hat F}_{\alpha}{\hat \rho}(t)\}
\label{f}
\end{equation}
to second order in $\{{\hat F}_{\beta}\}$, and to relate it to the motion of the centroid of 
the oscillator distribution. The centroid is the collection of mean displacements
$X_{\beta}(t) = {\rm Tr}\{{\hat X}_{\beta}{\hat \rho}(t)\}$, with velocities
$V_{\beta}(t) = {\dot X}_{\beta}(t) = {\rm Tr}\{{\hat P}_{\beta}{\hat \rho}(t)\}/M_{\beta}$
(not to be confused with the perturbing potential ${\hat V}$).
The centroid $\{ X_{\beta}(t) \}$ is what we would interpret as {\sl classical} coordinates, 
in a mixed quantum-classical picture of electron-nuclear dynamics, such as Ehrenfest dynamics \cite{tnt4}.

$F_{\alpha}(t)$ above is the mean force exerted by electrons on degree of freedom $\alpha$. 
Indeed
\begin{equation}
{\hat F}_{\alpha} = \frac{1}{{\rm i}\hbar}\,
\left[{\hat P}_{\alpha},\left(- \sum_{\beta}\,{\hat F}_{\beta}{\hat X}_{\beta}\right)\right]
\end{equation}
and therefore (as may be expected from the earlier discussion)
$F_{\alpha}(t)$ is the contribution to the mean rate of change of the momentum
of degree of freedom $\alpha$, due to the coupling to electrons \cite{sorbello}. 
In other words, $F_{\alpha}(t)$ is the effective
force experienced by the mean displacement $X_{\alpha}(t)$, and is what we would calculate
for the force on the corresponding classical degree of freedom in the Ehrenfest approximation.

For the unperturbed density matrix we now make the choice
${\hat \rho}_{0}(t) = {\hat \rho}_{0,e}(t)\otimes{\hat \rho}_{0,{\rm osc}}(t)$, where
${\hat \rho}_{0,{\rm osc}}(t)$ describes oscillators in time-evolving wavepackets with
\begin{eqnarray}
&&X_{\beta}(t) = {\rm Tr}_{{\rm osc}}\{{\hat X}_{\beta}{\hat \rho}_{0,{\rm osc}}(t)\}
= a_{\beta}\,\cos{(\omega_\beta t - \phi_{\beta})}\\
&&V_{\beta}(t) = {\dot X}_{\beta}(t) = - a_{\beta} \omega_\beta\,\sin{(\omega_\beta t - \phi_{\beta})}\,.
\end{eqnarray}
Below we consider non-interacting electrons. 
However screening plays an important - and occasionally problematic - role in current-induced 
forces and electron-phonon scattering. An outline is given in Refs. \cite{sorbello,tnt1}.

Next, we substitute into equation (\ref{1}), trace out the oscillators and trace out
all but one electron. The result, for non-interacting electrons, is
\begin{equation}
{\hat \rho}_{1e}(t) = {\hat \rho}_{0,1e}(t) - \frac{1}{{\rm i}\hbar}\,\sum_{\beta}\int_{0}^{t}\,
[{\hat f}_{\beta}(\tau - t),{\hat \rho}_{0,1e}(t)]X_{\beta}(\tau) \,d\tau
\label{2}
\end{equation}
where all operators are now 1-electron operators (indicated explicitly for the reduced 1-electron
DM and by the use of lower-case symbols for other electronic operators) and
\begin{eqnarray}
{\hat f}_{\beta}(s) =
{\rm e}^{{\rm i}{\hat h}_{e} s / \hbar}
{\hat f}_{\beta} {\rm e}^{-{\rm i}{\hat h}_{e} s / \hbar}
\end{eqnarray}
is an interaction-picture operator. \footnote{The connection with Ehrenfest dynamics is 
especially clear at this point: (\ref{2}) is what we would have written for the perturbed
electronic DM due to coupling to classical degrees of freedom with trajectories $\{ X_{\beta}(t) \}$;
the result, via (\ref{f}), are the corresponding forces in the Ehrenfest approximation.}

Our final task is the choice of unperturbed electronic DM ${\hat \rho}_{0,1e}(t)$.
To this end we now briefly introduce the Landauer picture of steady-state transport
in mesoscopic systems.

\subsection{Landauer picture of transport}

\begin{figure}[ht]
\begin{center}
\includegraphics[scale=0.40]{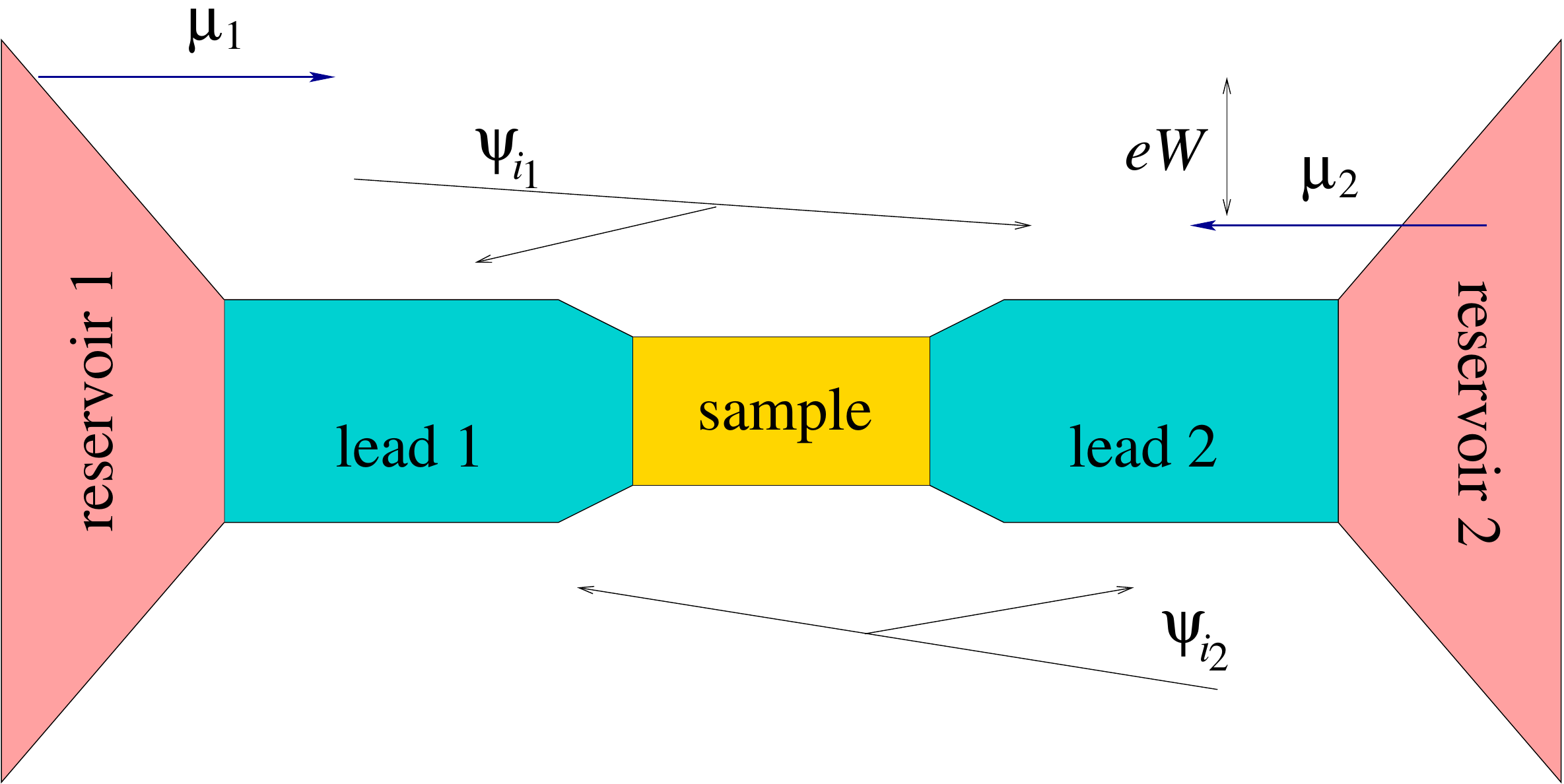}
\caption{\label{fig4} The Landauer picture. Details are discussed in the text.}
\end{center}
\end{figure}

The Landauer picture (LP) of conduction is a powerful construct which - combined with Green's function theory -
has become the most widely used transport method for nanoscale systems. The LP is summarized in Fig. \ref{fig4}.
The nanoconductor of interest - the sample - is connected to two perfect leads, each in turn
connected through a reflectionless contact to a heat-particle reservoir for electrons. The two
reservoirs have different electrochemical potentials, $\mu_1$ and $\mu_2$, and inject carriers into
their respective leads with the corresponding energy distributions. Formally, the injected carriers
in the steady state occupy a set of stationary scattering wavefunctions, two of which are depicted schematically
in the figure. A state $|\psi_{i_1}\rangle$ consists of an incident wave in lead 1 that gets partially transmitted
into lead 2 and partially reflected back, and conversely for $|\psi_{i_2}\rangle$.
Bias enters the LP via the {\sl electrochemical} potential difference $eW = \mu_1 - \mu_2$ between the
{\sl reservoirs}. \footnote{In principle the bias also modifies the Hamiltonian, due to the self-consistent
charge redistribution under current, but we will not consider this here.}

We may form the partial density-of-states operators for the two sets of states
\begin{equation}
{\hat d}_{1}(\epsilon) = \sum_{i_1}\,|\psi_{i_1}\rangle \delta(\epsilon - \epsilon_{i_1}) \langle \psi_{i_1}| \quad{\rm and}\quad
{\hat d}_{2}(\epsilon) = \sum_{i_2}\,|\psi_{i_2}\rangle \delta(\epsilon - \epsilon_{i_2}) \langle \psi_{i_2}|
\end{equation}
where $\epsilon_{i_1,i_2}$ are the energies of the respective states. Hence we can characterize the LP
steady state by the time-independent 1-electron DM
\begin{equation}
{\hat \rho}_{\rm LP} = \int_{-\infty}^{\infty} \, {\hat \rho}_{\rm LP}(\epsilon) \, d\epsilon \quad , \quad
{\hat \rho}_{\rm LP}(\epsilon) = n_F(\epsilon-\mu_1){\hat d}_{1}(\epsilon) + n_F(\epsilon-\mu_2) {\hat d}_{2}(\epsilon)
\end{equation}
where $n_F(\epsilon-\mu_1)$ and $n_F(\epsilon-\mu_2)$ are the Fermi-Dirac population functions for electrons 
arriving from the respective reservoirs, $n_F (\xi) = ({\rm e}^{\xi/k_{\rm B}T}+1)^{-1}$. 
Like the electrochemical potentials $\mu_1$ and $\mu_2$, the temperature $T$
therein is a property of the reservoirs. \footnote{We will work with spinless electrons. 
Spin may be included by adding a spin quantum number to 1-electron states.} 
${\hat d}_{1}(\epsilon)$ and ${\hat d}_{2}(\epsilon)$, and hence all 1-electron properties of the system, 
may be calculated with the aid of Green's functions \cite{tnt4}.

\subsection{Steady-state forces}

We now set ${\hat \rho}_{0,1e}(t) = {\hat \rho}_{\rm LP}$,
write $X_{\beta}(\tau) = a_{\beta}\,\cos{[\omega_\beta(\tau -t) + \omega_\beta t- \phi_{\beta}]}
= X_{\beta}(t) \,\cos{\omega_\beta(\tau -t)} +
[V_{\beta}(t)/\omega_\beta]\,\sin{\omega_\beta(\tau -t)}$, and substitute into (\ref{2}) and (\ref{f})
to obtain the second-order forces \footnote{Notice that within the chosen perturbation order,
in (\ref{force}) we are free to treat $\{ X_{\beta}(t) \}$ and $\{ V_{\beta}(t) \}$ as the
{\sl actual} - as opposed to unperturbed - oscillator positions and velocities.}
\begin{equation}
F_{\alpha}(t) = -\sum_{\beta}\,K_{\alpha\beta} X_{\beta}(t) + \sum_{\beta}\,L_{\alpha\beta} V_{\beta}(t)
\label{force}
\end{equation}
where
\begin{eqnarray}
&& K_{\alpha\beta} = \frac{1}{{\rm i}\hbar}\,\int_{0}^{t}\,{\rm Tr}\{ {\hat f}_{\alpha}
[{\hat f}_{\beta}(\tau - t),{\hat \rho}_{\rm LP}] \} \cos{\omega_\beta (\tau -t)}\,d\tau\, \label{kd}\\
&& L_{\alpha\beta} = - \frac{1}{{\rm i}\hbar \omega_\beta}\,\int_{0}^{t}\,{\rm Tr}\{ {\hat f}_{\alpha}
[{\hat f}_{\beta}(\tau - t),{\hat \rho}_{\rm LP}] \} \sin{\omega_\beta (\tau -t)}\,d\tau \,. \label{ld}
\end{eqnarray}
To evaluate $K_{\alpha\beta}$ and $L_{\alpha\beta}$, we take $t\rightarrow \infty$ \footnote{This limit
corresponds to a notional procedure in which we allow steady-state conditions to become reestablished after
the application of the electron-oscillator interaction.}
and use the relations
\begin{eqnarray}
&& {\rm e}^{{\rm i}{\hat h}_{e}s/\hbar}\,{\hat \rho}_{\rm LP}
= \int_{-\infty}^{\infty}\,{\rm e}^{{\rm i}\epsilon s/\hbar}\,{\hat \rho}_{\rm LP}(\epsilon)\,d\epsilon\\
&& {\hat g}^{+}(\epsilon) = \lim_{t\rightarrow \infty}\,
\frac{1}{{\rm i}\hbar}\,\int_{0}^{t}\,{\rm e}^{{\rm i}(\epsilon - {\hat h}_{e})s/\hbar}\,ds
\end{eqnarray}
where
\begin{equation}
{\hat g}^{+}(\epsilon) = \lim_{\eta\rightarrow 0^{+}}\,(\epsilon - {\hat h}_{e} + {\rm i}\hbar\eta)^{-1}
\end{equation}
is the retarded 1-electron Green's function. \footnote{A step that may be used in the evaluation of these limits
is to introduce a factor of ${\rm e}^{(\tau-t)\eta}$ in the integrand in (\ref{2}), describing
decoherence of the dynamics into the distant past.}

The result for $K_{\alpha\beta}$ and $L_{\alpha\beta}$ is
\begin{eqnarray}
K_{\alpha\beta} &=& {\rm Re}\,\int_{-\infty}^{\infty}\,
{\rm Tr} \{{\hat f}_{\alpha}[{\hat g}^{+}(\epsilon+\hbar\omega_\beta) +
{\hat g}^{+}(\epsilon-\hbar\omega_\beta)]{\hat f}_{\beta}{\hat \rho}_{\rm LP}(\epsilon)\}\,d\epsilon \label{K0} \\
L_{\alpha\beta} &=& \frac{1}{\omega_\beta}\,{\rm Im}\,\int_{-\infty}^{\infty}\,{\rm Tr} \{{\hat f}_{\alpha}
[{\hat g}^{+}(\epsilon+\hbar\omega_\beta) - {\hat g}^{+}(\epsilon-\hbar\omega_\beta)]
{\hat f}_{\beta}{\hat \rho}_{\rm LP}(\epsilon)\}\,d\epsilon\,.
\label{L0}
\end{eqnarray}
In the rest of the discussion of mean forces we will consider a single oscillator frequency, $\omega_0$.
An example where this situation arises are degenerate, or nearly degenerate, unperturbed normal modes, 
which are especially strongly coupled by current \cite{waterwheel,mads1}. 
Alternatively, we may think of the introduction of $\omega_0$ as follows. Once the full perturbed 
equations of motion of atoms under current are known, we may recalculate the normal modes to find new 
modes describing atomic motion in the current-carrying system \cite{mads1}. Due to the non-conservative
current-induced force, which we will discuss shortly, the amplitude of some of the new modes may grow
in time and dominate the dynamics.
$\omega_0$ can be the frequency (or more precisely its real part) of one of these modes, 
with indices $\alpha$ and $\beta$ labelling atomic degrees of freedom. The forces we are calculating are 
then the forces on atoms, under motion in that mode. The general case is discussed in \cite{langevin}.

\subsection{Electron-hole pairs}

Atomic motion generates electron-hole excitations in the electron gas which act back via the forces on atoms. 
It turns out that the forces can be written solely in terms of a weighted electron-hole density of states. 
We introduce the function $\Lambda_{\alpha\beta}(\omega)$, which encodes important information about electrons:
\begin{equation}
\Lambda_{\alpha\beta}(\omega) = \sum_{p,q} \, \langle \psi_q|\hat{f}_\alpha|\psi_p\rangle\langle
\psi_p|\hat{f}_\beta|\psi_q\rangle n_F(\epsilon_q-\mu_q)\delta(\hbar\omega-\epsilon_q+\epsilon_p)\,.
\end{equation}
Here the sums include all electron states, i.e. scattering states emerging both from lead 1 and from lead 2. 
We use the notation $\mu_q = \mu_1$ if the state $\psi_q$ is incident from lead 1, and likewise for lead 2.

The dynamics is controlled by two Green's functions, which only depend on the oscillator frequency and are 
independent of the details of the electronic system. They are
\begin{align}
G_K(\omega) &= \frac{1}{\omega+\omega_0 + {\rm i}\eta}+\frac{1}{\omega-\omega_0+{\rm i}\eta} \\
G_L(\omega) &= \frac{1}{\omega+\omega_0 + {\rm i}\eta}-\frac{1}{\omega-\omega_0+{\rm i}\eta}
\end{align}
where $\eta$ is an infinitesimal positive quantity. The two force-functions can now be written as
\begin{align}
K_{\alpha\beta} &= \mbox{Re}\,\int_{-\infty}^{\infty}\, G_K(\omega) \Lambda_{\alpha\beta}(\omega) \, d\omega \\
L_{\alpha\beta} &= \frac{1}{\omega_0}\,\mbox{Im}\,\int_{-\infty}^{\infty}\, G_L(\omega) \Lambda_{\alpha\beta}(\omega)
\,d\omega \,.
\end{align}
We can further symmetrize $\Lambda_{\alpha\beta}(\omega)$ and write it as a sum of even and odd functions:
\begin{equation}
\Lambda_{\alpha\beta}(\omega) = \Lambda^{\rm e}_{\alpha\beta}(\omega)+\Lambda^{\rm o}_{\alpha\beta}(\omega)
\end{equation}
where
\begin{equation}
\Lambda^{\rm e}_{\alpha\beta}(\omega) = 
\frac{1}{2}\,\left[ \Lambda_{\alpha\beta}(\omega)+\Lambda_{\alpha\beta}(-\omega)\right]\quad{\rm and}\quad
\Lambda^{\rm o}_{\alpha\beta}(\omega) = 
\frac{1}{2}\,\left[\Lambda_{\alpha\beta}(\omega)-\Lambda_{\alpha\beta}(-\omega)\right]\,.
\end{equation}
Since the real and imaginary parts of $G_K(\omega)$ and $G_L(\omega)$ are either even or odd, 
we arrive at the following forms for the functions (using the notation $A = A' + {\rm i}A''$ 
for the various complex quantities):
\begin{align}\label{forces}
K_{\alpha\beta} &= 2\,\int_0^\infty \, \left[ G'_K(\omega) {\Lambda_{\alpha\beta}^{\rm o}}'(\omega) - G_K''(\omega){\Lambda_{\alpha\beta}^{\rm e}}''(\omega)\right]\,d\omega\\
L_{\alpha\beta} &= 2\,\frac{1}{\omega_0}\,\int_0^\infty \, \left[ G''_L(\omega) {\Lambda_{\alpha\beta}^{\rm o}}'(\omega) + G_L'(\omega){\Lambda_{\alpha\beta}^{\rm e}}''(\omega) \right]\,d\omega
\end{align}
where the two relevant $\Lambda$-functions are
\begin{align}
{\Lambda_{\alpha\beta}^{\rm o}}'(\omega) = \frac{1}{2}\,&\sum_{p,q}\,\mbox{Re}\,\big[\langle \psi_q|\hat{f}_\alpha|\psi_p\rangle\langle \psi_p|\hat{f}_\beta|\psi_q\rangle\big] \times \nonumber\\
&\left[ n_F(\epsilon_q-\mu_q)-n_F(\epsilon_p-\mu_p)\right]\delta(\hbar\omega-\epsilon_q+\epsilon_p)\\
{\Lambda_{\alpha\beta}^{\rm e}}''(\omega) = \frac{1}{2}\,&\sum_{p,q}\,\mbox{Im}\,\big[\langle \psi_q|\hat{f}_\alpha|\psi_p\rangle\langle \psi_p|\hat{f}_\beta|\psi_q\rangle\big]\times \nonumber\\
&\left[ n_F(\epsilon_q-\mu_q)-n_F(\epsilon_p-\mu_p)\right]\delta(\hbar\omega-\epsilon_q+\epsilon_p)\,.
\end{align}

We note several points. First, both $\Lambda$-functions are a sum over electron-hole 
pairs with a total energy $\hbar\omega$. The factors 
$\left[ n_F(\epsilon_q-\mu_q)-n_F(\epsilon_p-\mu_p) \right]$, together with the delta function,
ensure that. Each electron-hole pair contribution is weighted by a factor depending on 
the matrix elements of the force operators. Next, by inspection we can see that the function 
${\Lambda_{\alpha\beta}^{{\rm o}}}'(\omega)$ is even with respect to exchange of indices $\alpha$ and $\beta$. 
The other function, ${\Lambda_{\alpha\beta}^{\rm e}}''(\omega)$, is odd in these indices.

The final, and very important, observation is that at equilibrium, i.e. $\mu_1 = \mu_2 = \mu$, 
the function ${\Lambda_{\alpha\beta}^{\rm e}}''(\omega)$ will vanish. Indeed then all
states (right- and left-travelling) with a given energy $\epsilon_q = \epsilon$ enter the expression for 
$\Lambda(\omega)$ with the same occupancy $n_F(\epsilon-\mu)$. But, within this degenerate Hilbert space, 
we can always choose a basis with pure real wavefunctions. Then ${\Lambda_{\alpha\beta}^{\rm e}}''(\omega)$, 
which is composed of the imaginary parts, vanishes.

\section{Position-dependent forces}

\subsection{General results}

Equation (\ref{force}) shows that matrix $K$ describes position-dependent steady-state forces
exerted by electrons on nuclear vibrations. They come in two forms: conservative and non-conservative,
which we separate below.

Forces are non-conservative if their ``curl'' is non-zero. The ``curl'' in our case is defined 
as the tensor
\begin{equation}
\frac{\partial F_{\beta}}{\partial X_{\alpha}} - \frac{\partial F_{\alpha}}{\partial X_{\beta}}
= K_{\alpha\beta}-K_{\beta\alpha}
\end{equation}
which is twice the anti-symmetric part of $K_{\alpha\beta}$,
\begin{equation}
K_{\alpha\beta}^{{\rm asym}} = -2\,\int_0^\infty \,
G_K''(\omega){\Lambda_{\alpha\beta}^{\rm e}}''(\omega)\,d\omega\,.
\end{equation}
The Green's function $G_K''(\omega)$ is given by
\begin{equation}
G_K''(\omega) = \mbox{Im} \, G_K(\omega) = - \pi \left[ \delta(\omega-\omega_0) + \delta(\omega+\omega_0) \right]\,.
\end{equation}
Hence our final result for the non-conservative force is
\begin{equation}
F^{\rm NC}_\alpha(t) = -2\pi\,\sum_\beta \,{\Lambda_{\alpha\beta}^{\rm e}}''(\omega_0) X_\beta(t)\,.
\end{equation}
The force will vanish at equilibrium, since as we have noted
the function ${\Lambda_{\alpha\beta}^{\rm e}}''(\omega_0)$ vanishes in that case.

The remaining - symmetric - part of $K_{\alpha\beta}$ on the other hand gives rise to
effective corrections to the equilibrium dynamical response matrix \cite{tnt3},
and thus to the effective confining potential for nuclei.
This correction is also potentially important, as it shifts phonon frequencies.

\subsection{Example: Sorbello's argument}

The anti-symmetric part of $K$ describes non-conservative forces with the capacity to drive the
nuclear subsystem around closed paths.
Insight into these forces may be gained by evaluating $K^{\rm asym}$ explicitly for a weak scatterer,
serving as a test particle, in a current-carrying metal \cite{tnt2}.
If the scattering potential of the test particle is
${\hat v} = C \delta({\hat {\bf r}} - {\bf R})$,
with ${\hat {\bf f}} = - {\bf \nabla}_{\bf R} {\hat v}$,
where ${\bf r}$ denotes electron position and ${\bf R}$ is that of the scatterer,
then to lowest order in the strength $C$ we get
\begin{equation}
{\bf \nabla}_{\bf R}\times {\bf F}({\bf R}) = \frac{2\pi m_e C^{2}}{\hbar}\,
{\bf \nabla}_{\bf R}\times {\bf L}({\bf R}) \,.
\label{dj}
\end{equation}
Here the vector ${\bf L}({\bf R})$ is given by
\begin{align}
{\bf L}({\bf R}) ={}& \frac{1}{2}\sum_{p,q}\,\big(\rho_p {\bf J}_q-\rho_q {\bf J}_p\big)\times \nonumber \\
&\left[ n_F(\epsilon_q-\mu_q)-n_F(\epsilon_p-\mu_p)\right]\delta(\hbar\omega_0-\epsilon_q+\epsilon_p)
\label{lcurl}
\end{align}
where $\rho_q = \psi_q^*({\bf R})\psi_q({\bf R})$ is the electron density of state 
$q$ at the test-particle position ${\bf r} = {\bf R}$, and 
${\bf J}_q = (\hbar/m_e)\, \mbox{Im}\,\psi_q^*({\bf R})\nabla \psi_q({\bf R})$ 
is the velocity density of the state, at that position.

We will evaluate it for the situation considered in Sorbello's
argument (Fig. \ref{fig2}). 
We consider a jellium wire with a square cross section of side $S$. 
The energy eigenstates are travelling waves along the wire axis ($z$) and 
standing waves in the transverse ($x$-$y$) plane:
\begin{equation}
\psi_{knm}({\bf r}) = \frac{{\rm e}^{ikz}}{\sqrt{l}}\,
\chi_{nm}({\bf r })\quad,\quad \chi_{nm}({\bf r}) =  
\frac{2}{S}\,\sin\left(\frac{n\pi x}{S}\right)\sin\left(\frac{m\pi y}{S} \right)
\end{equation}
where $l$ is a normalization length.
The energy of the state is $\epsilon_{knm} = \frac{\hbar^2 k^2}{2m_e}+ \frac{\hbar^2\pi^2}{2m_eS^2}(n^2+m^2)$. 

\begin{figure}
\begin{center}
\includegraphics[scale=1]{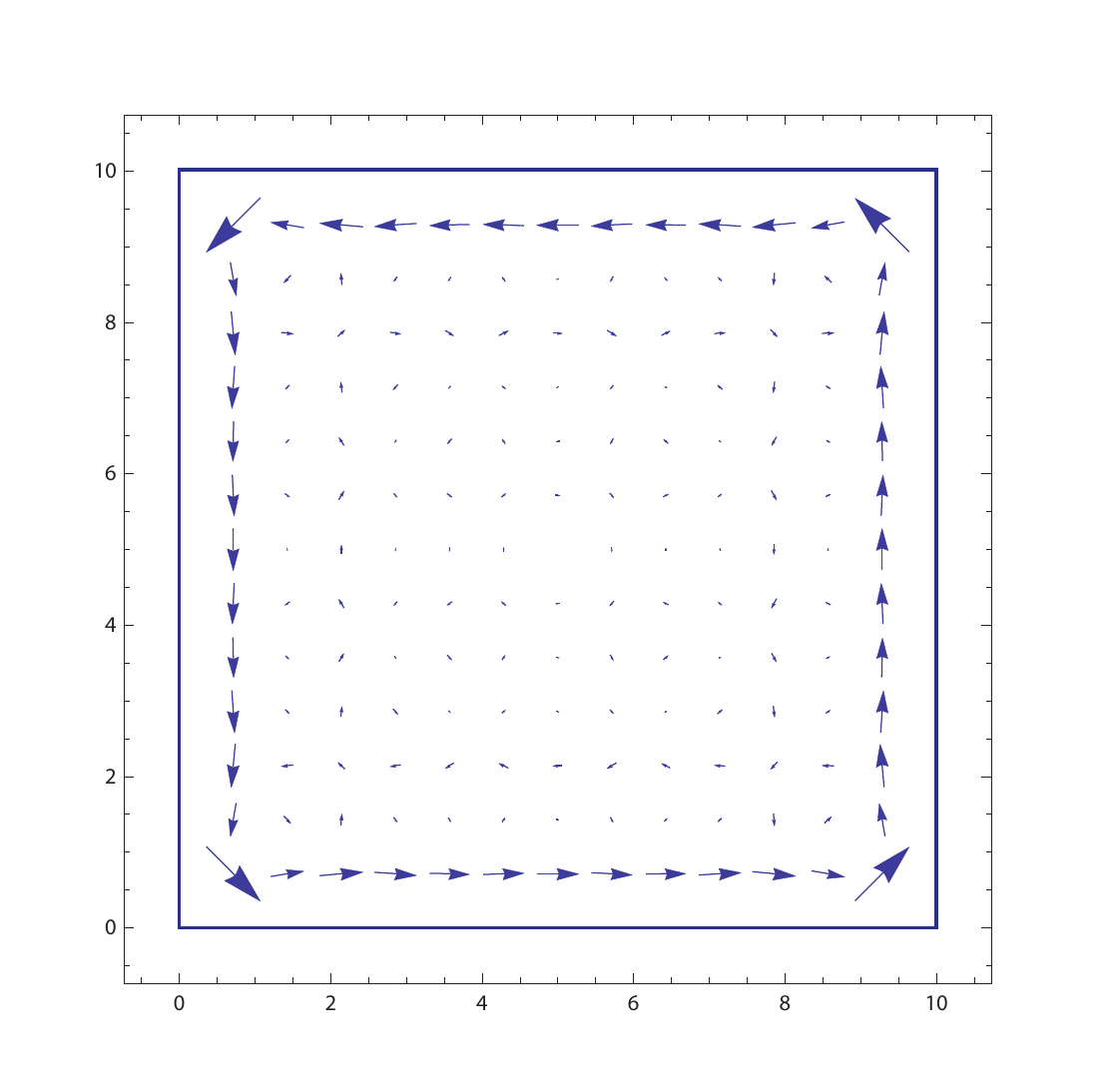}
\caption{\label{fig5} Curl of non-conservative force on an impurity in a 
current-carrying free-electron wire with a square cross section. Electrons are flowing out of the page.
The sidelength is $S = 10\lambda_F$.}
\end{center}
\end{figure}

\begin{figure}
\begin{center}
\includegraphics[scale=0.8]{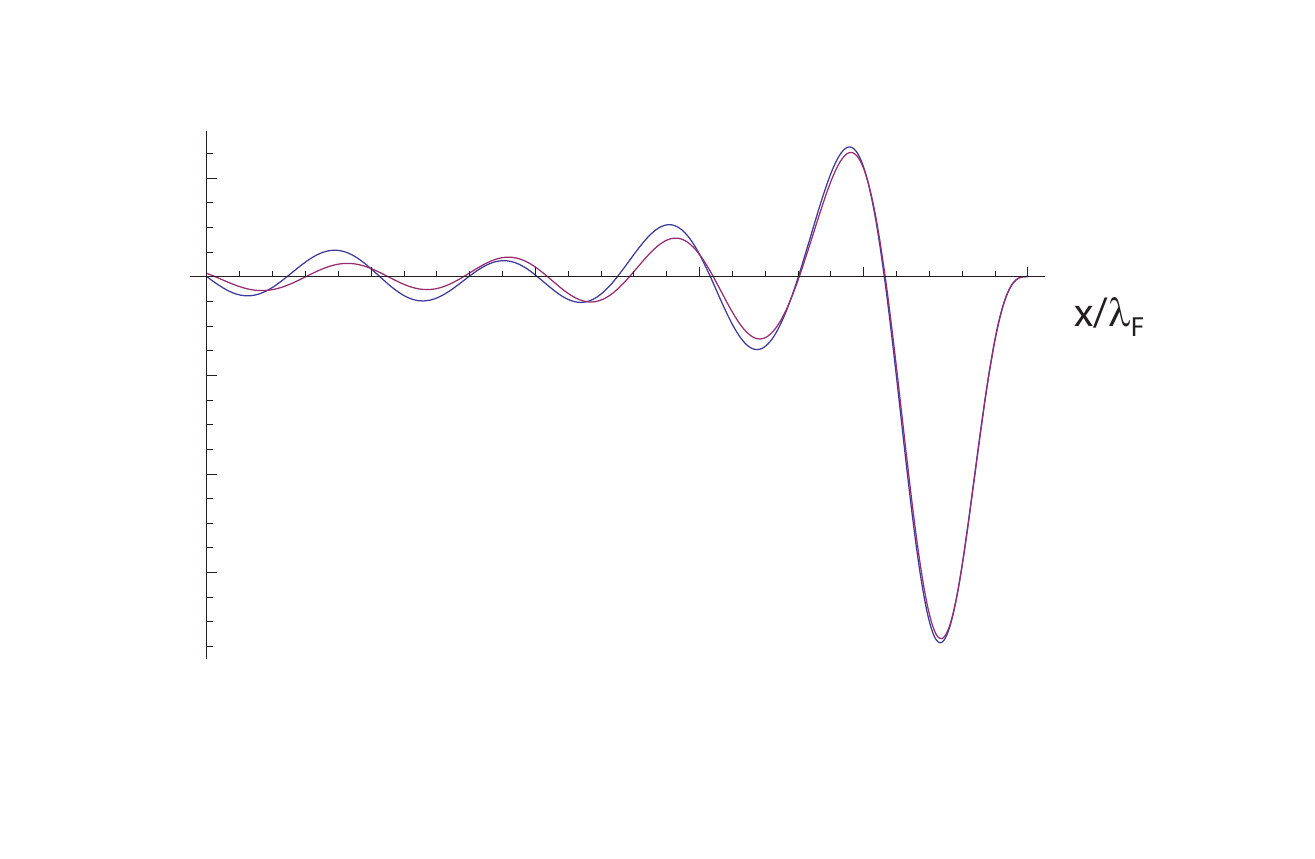}
\caption{\label{fig6} $y$-component of the curl of the force on the impurity close 
to the surface of the wire, for sidelengths $S = 10\lambda_F$ (blue) and $S= 20\lambda_F$ (red). 
Details are discussed in the text.}
\end{center}
\end{figure}

The curl of the non-conservative force will be proportional to $eW$ for small voltages. 
Figure \ref{fig5} shows a vector plot of the curl of the force, in the small-bias, small-frequency limit. 
We see that the curl is largest in a thin region of width $\lambda_F$ (the Fermi wavelength) 
close to the surface. 
Figure \ref{fig6} shows the $y$-component of the curl along a line parallel 
to the $x$-axis, through the wire centre, for the last 5 Fermi wavelengths close to the surface, 
for the cases $S=10 \lambda_F$ and $S=20 \lambda_F$. (The plot runs in the negative $x$-direction.)
We see that the curl of the non-conservative force is a surface phenomenon, which is independent 
of the size of the wire. Its origin is the variation in $\rho_q$ and ${\bf J}_q$ 
in equation (\ref{lcurl}) as we cross the wire surface. An analogous effect will be 
discussed in connection with another current-induced force below.

\section{Velocity-dependent forces}

\subsection{General results}
The velocity-dependent forces are controlled by the function $L_{\alpha\beta}$. 
First, we consider the symmetric part. It is given by
\begin{equation}
L_{\alpha\beta}^{\rm sym} = 2\,\frac{1}{\omega_0}\,\int_0^\infty \, 
G''_L(\omega) {\Lambda_{\alpha\beta}^{\rm o}}'(\omega)\,d\omega\,.
\end{equation}
The function $G''_L(\omega)$ is given by
\begin{equation}
G''_L(\omega) = \mbox{Im} \,G_L(\omega) = \pi \left[ \delta(\omega-\omega_0) - \delta(\omega+\omega_0)\right]\,.
\end{equation}
Hence this velocity-dependent force is
\begin{equation}
F^{\rm F}_\alpha(t) = \frac{2\pi}{\omega_0}\, \sum_\beta\, {\Lambda_{\alpha\beta}^{\rm o}}'(\omega_0) V_\beta(t)\,.
\end{equation}
This force can do work on the oscillating atoms. The power is
\begin{equation}
w = \sum_\alpha\, F^{\rm F}_\alpha V_{\alpha} 
= \frac{2\pi}{\omega_0}\,\sum_{\alpha,\beta} \,{\Lambda_{\alpha\beta}^{\rm o}}'(\omega_0) V_{\alpha}V_{\beta}
\end{equation}
which is non-vanishing in general. For most situations the power is negative and energy is transferred from vibrations 
to electrons. Hence the superscript ${\rm F}$ for friction. One can however have systems out of equilibrium
where the power in fact is positive \cite{negfric}.

Turning next to the anti-symmetric part, it is given by
\begin{equation}\label{Lasym}
L_{\alpha\beta}^{\rm asym} = 
2\,\frac{1}{\omega_0}\,\int_0^\infty \,G_L'(\omega){\Lambda_{\alpha\beta}^{\rm e}}''(\omega) \, d\omega\,.
\end{equation}
An anti-symmetric force
\begin{equation}\label{LorentzForce}
F^{\rm B}_\alpha = \sum_{\beta}\, L^{\rm asym}_{\alpha\beta} V_\beta
\end{equation}
will do no net work, since the power is
\begin{equation}
w = \sum_{\alpha,\beta}\, L^{\rm asym}_{\alpha\beta} V_\alpha V_\beta = 0 \,.
\end{equation}
This is a force akin to the Lorentz force. 
It can only bend trajectories. 
It can be related to the Berry phase of electrons 
as the atoms are moving around in the electron gas \cite{mads1,mthomas,berry}, 
and hence the superscript ${\rm B}$.

The function $G_L'(\omega)$ is even in $\omega$ and is given by
\begin{equation}
G_L'(\omega) = \mbox{Re}\,G_L(\omega) 
= \frac{\omega+\omega_0}{(\omega+\omega_0)^2+\eta^2}-\frac{\omega-\omega_0}{(\omega-\omega_0)^2+\eta^2}\,.
\end{equation}
This function decays on the frequency scale $\omega_0$, and integrates to zero.
Hence the integral in (\ref{Lasym}) is measuring variations in the electron-hole 
density of states ${\Lambda_{\alpha\beta}^{\rm e}}''(\omega)$ 
on the scale $\omega_0$. $F^{\rm B}$ will therefore be sensitive to the frequency 
structure of the electron-hole density of states,
and to features, such as resonances, on the scale of $\omega_0$ in particular. 
A flat electron-hole density of states will produce a vanishing ``Lorentz'' force.

\subsection{Example: friction}

Let us evaluate the electronic friction on the weak test scatterer considered earlier,
now in an equilibrium homogenous electron gas for simplicity. We need to evaluate the 
function ${\Lambda_{\alpha\beta}^{{\rm o}}}'(\omega_0)$. 
The friction will be isotropic, so we only need the case $\alpha=\beta=Z$. 
The result in the small-frequency limit is

\begin{equation}
L_{ZZ} = - p_{F} \sigma \rho
\end{equation}
where $\rho$ is the electron number density. The friction force when the scatterer is moving with velocity
${\bf V}$ therefore is ${\bf F}^{\rm F} = - p_{F} \sigma \rho {\bf V}$. But physically this situation is the same
as if the electron gas were moving past a stationary scatterer, at drift velocity $-{\bf V}$. Then the wind force from
(\ref{fw}) is the same as ${\bf F}^{\rm F}$, after we identify $ \rho (-{\bf V}) $ as the current density
${\bf j}$ in the rest frame of the scatterer.

Therefore we arrive at the expected but useful result that for a free-electron metal the friction and
the wind force are reciprocals of each other. This result is connected with the relation between the
friction and resistivity \cite{persson}, and between resistivity and electromigration forces \cite{landauer}.

\subsection{Example: ``Lorentz'' force}

We have seen that a frequency-independent function 
${\Lambda_{\alpha\beta}^{\rm e}}''(\omega)$ will not result in a Lorentz-like force. 
We consider our prime example of a particle moving in a flow of electrons in a thin jellium wire. 
It can be thought of as a set of 1d wires (or channels) - one for each transverse electron mode. 
Each channel will have a density of states with a van Hove singularity at the bottom of the respective subband. 
These strong variations of the density of states may provide the frequency dependence of 
${\Lambda_{\alpha\beta}^{\rm e}}''(\omega)$ that is needed.

We consider again a test scatterer located somewhere in the square cross section of the 
wire, used earlier. According to the general formula (\ref{LorentzForce}), 
the defect will feel a ``Lorentz'' force, ${\bf V}\times {\bf B}$, where the ``B''-field 
has a vector potential given by
\begin{align}
{\bf A}({\bf R}) ={}& \frac{m_e C^2}{2\hbar\omega_0}\,\int_0^\infty \,
G'_L(\omega) \,\sum_{p,p'}\,\big(\rho_{p'} {\bf J}_p-\rho_p {\bf J}_{p'}\big)\times \nonumber \\
&\left[ n_F(\epsilon_p-\mu_p)-n_F(\epsilon_{p'}-\mu_{p'})\right]\delta(\hbar\omega-\epsilon_p+\epsilon_{p'})\,
d\omega\,.
\end{align}
The 1-electron state $p$ has three quantum numbers $k,\,n,\,m$, 
where $k$ is the momentum along the wire, while $n,\,m$ are the transverse-mode quantum numbers,
with corresponding mode wavefunctions $\chi_{nm}$.
It is useful to express energies in terms of the transverse energy scale
$\epsilon_t = \frac{\hbar^2\pi^2}{2m_e S^2}$. 
The Fermi energy is then $\epsilon_F/\epsilon_t = (2S/\lambda_F)^2$. 
Likewise, the electron energies are $\epsilon_{knm}/\epsilon_t = \kappa^2 + n^2+m^2$, where $\kappa = kS/\pi$. 

The vector potential can then be written as 
\begin{align}
{\bf A}({\bf R}) ={}& \frac{C^2}{\hbar\omega_0}\,\frac{\pi}{4S^{3}} \,
\sum_{n,m,n',m'}\, \chi_{nm}^{2}({\bf R})\chi_{n'm'}^{2}({\bf R}) \times\nonumber \\
& \int_0^\infty d\kappa \int_0^\infty d\kappa' \, G'_L\left[\epsilon_t(\kappa^2 + n^2+m^2-\kappa'^2-n'^2-m'^2)/\hbar\right] \nonumber\\
& \kappa\, \left[n_F(\epsilon_{knm}-\mu_1)-n_F(\epsilon_{knm}-\mu_2)\right]\,{\bf e}_Z
\end{align}
where ${\bf e}_Z$ is the unit vector along the wire axis.

The integrals can in fact be done analytically at zero temperature.
The result is
\begin{align}
{\bf A}({\bf R}) ={}& \frac{C^2}{\hbar\omega_0}\,\frac{\pi}{4S^{3}} \,\frac{\pi\hbar}{2\epsilon_t^{3/2}}
\sum_{n,m,n',m'}\, \chi_{nm}^{2}({\bf R})\chi_{n'm'}^{2}({\bf R}) \times\nonumber \\
&\mbox{Re}\,\bigg(
\sqrt{(n'^2+m'^2)\epsilon_t-\mu_1-\hbar\omega_0}
-\sqrt{(n'^2+m'^2)\epsilon_t-\mu_2-\hbar\omega_0} \nonumber \\
&-\sqrt{(n'^2+m'^2)\epsilon_t-\mu_1+\hbar\omega_0}
+\sqrt{(n'^2+m'^2)\epsilon_t-\mu_2+\hbar\omega_0}\bigg)\,{\bf e}_Z\,.
\end{align}
Here the sum over $n,\,m$ is restricted to values such that $\epsilon_t(n^2+m^2)<\mu_2$,
and we have assumed that no subband edges fall within the bias window.

This result enables the following observations.
First, under lateral confinement $\chi_{nm}({\bf R})$ are standing waves.
The resultant interference ripples as a function of transverse position in the wire
generate a non-vanishing effective ``magnetic'' field, even in the absence
of longitudinal inhomogeneities or backscattering in the conductor.

Consider next the gross structure of the vector potential. To this end suppose
that we average over the ripples due to the confinement, 
and treat each $\chi_{nm}({\bf R})$ as a constant inside the wire, dropping to zero outside.
This still leaves the dramatic variation in ${\bf A}$ as we cross the wire surface, 
giving rise to a surface ``magnetic'' field, analogous to the curl of the non-conservative 
force discussed earlier.

Therefore the ``Lorentz'' forces are of particular interest for surface atoms and adsorbates 
in metallic nanowires, with departures from the dynamics that might be expected otherwise.

We conclude this section with a comment on the range of validity of the present discussion.
The perturbative approach above requires restricted mean displacements over long times. This
precludes free translation of scattering centres as a type of motion we can consider. In fact,
if we allow a maximum typical displacement $a_{\rm max}$, then for motion at typical velocity $V$
the present approach requires a minimum frequency $\omega_{\rm min} \sim V/a_{\rm max}$.
Conversely, for a given oscillator frequency $\omega$, we require velocities below 
$V_{\rm max} \sim a_{\rm max} \omega$. An important avenue - in this context and 
more generally - is the expansion of the Hamiltonian in (\ref{heph})
to higher order in the oscillator displacements. 

Finally, we may describe the ``Lorentz'' force as a dynamical
effect in the following sense. Keeping all else fixed, the ratio of the current-induced 
velocity-dependent ``Lorentz'' force to the current-induced displacement-dependent non-conservative force scales 
with frequency as $\hbar \omega / \epsilon_e$, where $\epsilon_e$ is a pertinent electronic energy 
scale \cite{langevin}. Therefore in the small-frequency limit the static non-conservative force dominates.

\section{Force noise}

As discussed earlier, $F_{\alpha}(t)$ is the force that enters the so-called Ehrenfest approximation.
Physically, this is the mean force on nuclei, which does not yet tell us anything about force {\sl noise}.

We may immediately infer a key limitation of the Ehrenfest
approximation. Consider a problem with equilibrium electrons at an elevated electronic temperature.
Then both the non-conservative and Lorentz-like force vanish, and we are left with motion under
conservative forces, along with the friction, leading to an inevitable loss of
energy from the nuclear motion, whatever the nuclear kinetic energies.

But this cannot be right: hot enough electrons should be delivering energy to the atomic
motion, not the other way round.

We conclude that there must be a key physical process missing from the Ehrenfest approximation.
Indeed, this is {\sl spontaneous phonon emission} \cite{kab,theilhaber,disscomp} - the primary way
for excited electrons to excite atomic vibrations. We conclude further that the missing force noise
must be the agent responsible for spontaneous emission and ultimately for electron-phonon
thermal equilibration.

We can recover this noise by
considering the second-order corrections to the equation of motion 
of the reduced oscillator DM, ${\hat \rho}_{\rm osc}(t) = {\rm Tr}_e\{ {\hat \rho}(t) \}$.

\subsection{General results}

To this end we place (\ref{1}) into the many-body Liouville equation
\begin{equation}
{\rm i}\hbar \, {\dot {\hat \rho}}(t) = [{\hat H}_{0},{\hat \rho}(t)]
- \sum_{\beta}\,[{\hat F}_{\beta}{\hat X}_{\beta},{\hat \rho}(t)]
\end{equation}
and trace out electrons. This gives
\begin{equation}
{\rm i}\hbar \, {\dot {\hat \rho}}_{\rm osc} (t) =
[{\hat H}_{\rm osc},{\hat \rho}_{\rm osc}(t)] + {\hat \Xi}_1(t) + {\hat \Xi}_2(t)
\label{rhoscdot}
\end{equation}
where ${\hat H}_{\rm osc}$ is the harmonic oscillator Hamiltonian and the first- and second-order corrections
are given by
\begin{eqnarray}
{\hat \Xi}_1(t) &=& - \sum_{\beta} \, [{\hat X}_{\beta},{\hat \rho}_{0,{\rm osc}}(t)]\,F_{\beta}^{(0)}(t) \\
{\hat \Xi}_2(t) &=& - \sum_{\beta} \, [{\hat X}_{\beta}, {\hat \Phi}_{\beta}(t) ] \label{Xi2}
\end{eqnarray}
with $F_{\beta}^{(0)}(t) = {\rm Tr}\{{\hat F}_{\beta}{\hat \rho}_{0}(t)\}
= {\rm Tr}_{e}\{{\hat F}_{\beta}{\hat \rho}_{0,e}(t)\}$ and
\begin{eqnarray}
{\hat \Phi}_{\beta}(t)  = &-& \frac{1}{2{\rm i}\hbar}\,\sum_{\alpha}\int_{0}^{t}
\,[{\hat X}_{\alpha}(\tau - t),{\hat \rho}_{\rm osc}(t)]\,
\langle \{ {\hat F}_{\beta},{\hat F}_{\alpha}(\tau - t) \} \rangle_{t} \,d\tau \nonumber \\
&-& \frac{1}{2{\rm i}\hbar}\,\sum_{\alpha}\int_{0}^{t}\,
\{{\hat X}_{\alpha}(\tau - t),{\hat \rho}_{\rm osc}(t)\}\,
\langle [ {\hat F}_{\beta},{\hat F}_{\alpha}(\tau - t) ] \rangle_{t} \,d\tau\,
\label{corr} \\
= &-& \frac{1}{2{\rm i}\hbar}\,\sum_{\alpha}\int_{0}^{t}
\,[{\hat X}_{\alpha}(\tau - t),{\hat \rho}_{\rm osc}(t)]\,
\langle \{ {\hat F}_{\beta},{\hat F}_{\alpha}(\tau - t) \} \rangle_{t} \,d\tau \nonumber \\
&-& \frac{1}{2{\rm i}\hbar}\,\sum_{\alpha}\int_{0}^{t}\,
\{{\hat X}_{\alpha}(\tau - t),{\hat \rho}_{\rm osc}(t)\}\,
\langle [ {\hat f}_{\beta},{\hat f}_{\alpha}(\tau - t) ] \rangle_{t} \,d\tau\,.
\label{corr1}
\end{eqnarray}
Errors are of order 3 or higher in the force operators;
${\hat X}_{\alpha}(s)$ is an interaction-picture operator;
$\langle \dots \rangle_t$ denotes averaging in the instantaneous unperturbed DM at time $t$;
commutators are directly convertible to 1-electron form and hence (\ref{corr1}).

Term 1 in (\ref{corr1}), together with ${\hat \Xi}_1(t)$,
corresponds to a fictitious perturbation
$\displaystyle{- \sum_{\beta} {\hat X}_{\beta} f_{\beta}(t)}$, 
with $f_{\beta}(t)$ an {\sl effective} classical random force with mean 
$\langle  f_{\beta}(t) \rangle = F_{\beta}^{(0)}(t)$ and with correlation function
\begin{eqnarray}
\langle f_{\beta}(t)f_{\alpha}(\tau) \rangle
= \frac{1}{2}\,\langle \{ {\hat F}_{\beta},{\hat F}_{\alpha}(\tau - t) \} \rangle_{t}
= \frac{1}{2}\,\langle \{ {\hat F}_{\beta}(t),{\hat F}_{\alpha}(\tau) \} \rangle_{t=0}
\label{corrf}
\end{eqnarray}
after averaging over force realizations (denoted by $\langle \dots \rangle$ above).

Now we show that term 2 in (\ref{corr1}) corresponds to the forces found earlier. 
To this end we investigate the Wigner function
\begin{equation}
{\tilde \rho}_{\rm osc}({\bf X},{\bf P},t) = \frac{1}{(2\pi\hbar)^{N}}\,\int\,
\langle {\bf X}-{\bf S}/2| {\hat \rho}_{\rm osc}(t) |  {\bf X}+{\bf S}/2 \rangle \,
{\rm e}^{{\rm i}{\bf P}\cdot{\bf S}/\hbar}\,d{\bf S}
\end{equation}
for $N$ oscillators. (Here ${\bf X}=(X_1,X_2,\dots,X_N)$ for short,
and similarly for ${\bf S}$ and ${\bf P}$.)

Consider the contribution of term 2 to ${\dot {\tilde \rho}}_{\rm osc}({\bf X},{\bf P},t)$.
Placing the relation
\begin{equation}
{\hat X}_{\alpha}(s) = {\hat X}_{\alpha} \cos{(\omega_{\alpha} s)}
+ \frac{{\hat P}_{\alpha}}{M_{\alpha}\omega_{\alpha}}\,\sin{(\omega_{\alpha} s)}
\end{equation}
in term 2 in (\ref{corr1}), comparing the integrals with (\ref{kd}) and (\ref{ld}),
and substituting back into (\ref{Xi2}) and thence into (\ref{rhoscdot}), 
we see that the quantity to be Wigner-transformed is
\begin{equation}
{\hat Q} = \frac{1}{2{\rm i}\hbar}\,\sum_{\alpha,\beta}\,
\left[ {\hat X}_{\beta}, \{ (K_{\beta\alpha}{\hat X}_{\alpha} - L_{\beta\alpha}{\hat P}_{\alpha}/M_{\alpha}),{\hat \rho}_{\rm osc}(t) \} \right]\,.
\end{equation}
The result is
\begin{equation}
{\tilde Q} + \sum_{\alpha} \,\frac{L_{\alpha\alpha}}{M_{\alpha}}\,{\tilde \rho}_{\rm osc}
= \sum_{\alpha,\beta}\,\frac{ \partial {\tilde \rho}_{\rm osc} }{ \partial P_{\beta} } K_{\beta\alpha}X_{\alpha}
- \sum_{\alpha,\beta}\,\frac{ \partial {\tilde \rho}_{\rm osc} }{ \partial P_{\beta} }
L_{\beta\alpha}\,\frac{P_{\alpha}}{M_\alpha} \,.
\label{qtilde}
\end{equation}
Upon comparing with the classical Liouville equation, obeyed exactly by the harmonic-oscillator
Wigner function,
\begin{equation}
{\dot {\tilde \rho}}_{\rm osc} = - \sum_{\beta} \, V_{\beta} \frac{ \partial  {\tilde \rho}_{\rm osc}}{ \partial X_{\beta} }
- \sum_{\beta} \, {\cal F}_{\beta} \frac{ \partial {\tilde \rho}_{\rm osc} }{ \partial P_{\beta} }
\end{equation}
where ${\cal F}_{\beta}$ is the total force on degree of freedom $\beta$, we recover (\ref{force}).

The extra term on the left in equation (\ref{qtilde}) is the known correction to
the Liouville equation required to conserve phase-space probability in the presence of dissipation
(in this case, the electronic friction). We note further that care is required in identifying
$P_{\alpha}/M_{\alpha}$ as $V_{\alpha}$ above, as the two quantities are in general not the same.
However any difference between them in the present case involves the force operators, and
therefore - within the given order of perturbation theory - the identification is justified.

The properties of the force noise and resultant phenomena, such as the thermal equilibration of
vibrations with the electron bath and Joule heating in nanowires under current, are discussed
in detail in Ref. \cite{langevin}.

\subsection{Example: thermal equilibration}

We consider a single oscillator coupled to an equilibrium electron bath.
The only forces present are the harmonic restoring force, the friction and the noise. We will
show that the competition between the latter two enables the oscillator to equilibrate with the bath.

First we must work out the correlation function (\ref{corrf}). 
The result for non-interacting independent electrons is
\begin{equation}
\langle f(t)f(\tau) \rangle = c(\tau-t)\quad , \quad
c(s) =  {F^{(0)}}^2 + \sum_{p,q}\,f_{pq}f_{qp}\,n_{p}(1-n_{q})\,\cos{(\omega_{pq}s)}
\end{equation}
where as before the indices label 1-electron states with energies $\{ \epsilon_p \}$. 
The occupancies $\{ n_p \}$ are given by the Fermi-Dirac distribution, $n_p = n_{F}(\epsilon_p - \mu)$,
and $\hbar \omega_{pq} = \epsilon_p - \epsilon_q$.

We now show that a {\sl classical} oscillator, under the above random force and the friction,
equilibrates at mean energies that in fact obey the quantum-mechanical Bose-Einstein distribution.
A harmonic oscillator of mass $M$ and angular frequency $\omega_0$, subjected to an additional
external force $f(t)$, undergoes velocity deviations from its native trajectory given by
\begin{equation}
\Delta V(t) = \frac{1}{M}\,\int_{0}^{t}\,f(s)\cos{\omega_0 (s - t)}\,ds\,.
\end{equation}
If $f(t)$ is the random force above, then the power at time $t$, 
after averaging and taking the long-time limit, is
\begin{equation}
w = \langle f(t) \Delta V(t) \rangle = \frac{1}{M}\,\int_{-\infty}^{0}\,c(s) \cos{(\omega_0 s)}\,ds\,.
\end{equation}
After some algebra this becomes
\begin{equation}
w = \frac{\pi\hbar}{2M}\,\coth{(\hbar\omega_0 /2k_{\rm B}T )}\,\sum_{p,q}\,f_{pq}f_{qp}\,(n_p - n_q)\,
\delta(\epsilon_p - \epsilon_q + \hbar \omega_0 )
\end{equation}
where $T$ is the temperature of the electron bath.

We now compare with the friction coefficient from (\ref{L0}), producing the relation
\begin{equation}
w = \frac{\hbar\omega_0}{2M}\,|L|\,\coth{(\hbar\omega_0 /2k_{\rm B}T)}
\end{equation}
which is an expression of the Fluctuation-Dissipation Theorem.

Setting $w$ equal to the power lost to the friction, $|L| \langle V^2 \rangle$,
produces the steady-state oscillator energy
\begin{equation}
\langle E_{\rm osc}\rangle = M\langle V^2 \rangle =
\hbar\omega_0\,\left( \frac{1}{2} + \frac{1}{{\rm e}^{\hbar\omega_0/k_{\rm B}T} - 1}\right)
\end{equation}
which is the Bose-Einstein distribution, for the given bath temperature.

Therefore the {\sl effective} classical random force that was {\sl identified} within a
quantum-mechanical calculation, when applied together with the quantum-mechanical
friction to a fictitious classical oscillator, has exactly the properties required to
reproduce the behaviour expected from the full interacting quantum problem.
One of these properties is the zero-point oscillator energy that survives even in the limit
of zero bath temperature. These conclusions furthermore remain valid
under non-equilibrium conditions, resulting in a remarkable mapping of quantum phonons
coupled to an electron bath onto a classical stochastic problem \cite{langevin}.

\section{Summary}

Electron-nuclear dynamics in atomic-scale conductors is a challenging problem, combining
many-body physics with the description of open quantum systems driven out of equilibrium. 
A further difficulty, at the root of some of the controversies that the field has 
experienced, is the conceptual question of what we mean by force under these general conditions 
and, accordingly, what framework is required for its calculation.

The aim of this article is to provide an introduction to elements of the current
methodological and physical understanding of forces on nuclei in conducting systems. 
Two of the five forces - the effective conservative confining potential and the electronic 
friction - are present at equilibrium (though they collect non-equilibrium corrections). 
So too is the effective random force, responsible for Joule heating and electron-phonon 
equilibration. The non-conservative and ``Lorentz'' force are only possible under current.

We have aimed at obtaining these forces in a simple way, while illustrating 
the physics behind them with examples. The analogy with other flows however can be deceptive.
The current-induced force on a single defect in jellium (which is always along the incident
particle current) {\sl is} simple, and may be understood directly in terms of momentum transfer 
on the defect. But for nuclei in a solid current-induced forces can be counter-intuitive, 
because electrons then undergo multiple scattering in the entire region around a chosen target, 
with quantum-mechanical interference. This can result in forces on individual atoms with 
unexpected directions, including situations when the current-induced force on a defect 
opposes the incident electron ``beam''.

However the possibility of directional generalized angular-momentum transfer to the atomic motion
is a general property of current-carrying nanowires, resulting in a potent energy-transfer mechanism.
There are several active lines of research into these forces at present: the interplay between the 
stochastic and the non-conservative force, the properties of these forces in resonant systems 
(with sharp energy features), the capacity of the non-conservative forces to drive electromigration 
or mechanical failure. There continually is scope for further fundamental and conceptual work in the 
area, with a potential practical pay-off, such as the question of whether - and how - one might 
go about subsuming current-induced forces into effective corrections to semi-empirical interatomic 
potentials.

It is hoped that the present discussion will be of assistance to the general reader in following up
past and future work in this field.

\section*{Acknowledgements}
TNT and DD gratefully acknowledge funding from EPSRC (EP/I00713X/1).
JTL acknowledges support from NSFC (no. 61371015 and 11304107).

\end{document}